\pdfoutput=1
\documentclass{ws-mpla}
\usepackage[super]{cite}
\usepackage{xcolor,dsfont}
\usepackage[verbose,hypertexnames=false,
      colorlinks=true,
      linkcolor=blue,
      urlcolor=cyan,
      filecolor=magenta,
      citecolor=red,
      pdfstartview=FitV,
      hyperfootnotes=false,
      pdftitle={Carroll spinors},
        pdfauthor={Daniel Grumiller, Lea Mele, Luciano Montecchio},
        pdfsubject={Carroll spinors},
        pdfkeywords={Carroll, spinors, ELKO},
        bookmarksopen=true
      ]{hyperref}
\hypersetup{colorlinks=true,allbordercolors=blue,pdfborderstyle={/S/U/W 1}}

\DeclareMathOperator{\extdm}{d}
\newcommand{\extd}{\extdm \!}

\newcommand{\eq}[2]{\begin{equation} #1 \label{#2} \end{equation}}

\begin{document}


\newcommand{\mytitle}{Carroll spinors}
\markboth{Daniel Grumiller, Lea Mele, and Luciano Montecchio}{\mytitle}

\catchline{}{}{}{}{}

\title{\mytitle}

\author{Daniel Grumiller}

\address{Institute for Theoretical Physics, TU Wien\\ 
Wiedner Hauptstrasse 8-10, A-1040 Vienna, Austria, Europe\\
\href{mailto:grumil@hep.itp.tuwien.ac.at}{\texttt{grumil@hep.itp.tuwien.ac.at}}}

\author{Lea Mele}

\address{Service de Physique de l’Univers, Champs et Gravitation, Universit\'e de Mons -- UMONS\\
20, place du Parc, B-7000 Mons, Belgium\\
\href{mailto:Lea.Mele@umons.ac.be}{\texttt{Lea.Mele@umons.ac.be}}
}

\author{Luciano Montecchio}
\address{Institute for Theoretical Physics, TU Wien\\ 
Wiedner Hauptstrasse 8-10, A-1040 Vienna, Austria, Europe\\
\href{mailto:luciano.montecchio@tuwien.ac.at}{\texttt{luciano.montecchio@tuwien.ac.at}}}

\maketitle

\begin{history}
\received{(30 October 2025)}
\end{history}

\begin{abstract}
This crisp summary of salient aspects of Carroll spinors is dedicated to the memory of Dharam Ahluwalia, the intrepid champion of ELKO spinors.
\end{abstract}


\ccode{PACS Nos.: 03.65.Pm, 11.30.Cp, 11.10.-z, 04.20.Cv}

\paragraph{Opening remark.} Paragraphs in italics are brief reminiscences by the first author.


\section{Spinors}

\textit{Since this memorial volume is dedicated to Dharam Ahluwalia's scientific legacy, I find the most appropriate topic is spinors. Dharam was fascinated by spinors. Perhaps the best evidence for this is that eleven of his twelve best-known papers prominently involve spinors, either neutrinos \cite{Ahluwalia:1994uy,Ahluwalia:1996ev,Ahluwalia:1998xb,Stancu:1999ct,Ahluwalia:2000iw} or ELKO \cite{Ahluwalia:2004sz,Ahluwalia:2004ab,Ahluwalia:2008xi,Ahluwalia:2009rh,Ahluwalia:2010zn} and their precursor \cite{Ahluwalia:1993zt}. Even though I never elaborated in scientific publications\cite{daRocha:2005ti,Boehmer:2007ut,daRocha:2007pz,Boehmer:2008rz,daRocha:2008we,Boehmer:2010ma,daRocha:2011yr,Wei:2011yr,Cavalcanti:2014wia,Bahamonde:2017ize,Ahluwalia:2022ttu} on our joint ELKO work, it was a pleasure to interact with Dharam in 2003-2005; a measure for the intensity is the number of e-mails we exchanged, over 500 in this period.} 

In this paper, we summarize recent results on spinors in a context that seems exotic enough that Dharam might have appreciated its original value --- and yet, this topic has started to invade the mainstream, due to several unexpected connections, some of which we will highlight in the conclusions. The main protagonists of this work are Carroll spinors, and its main goal is to explain what they are and why we should care about them.

Before doing so, recall that the Clifford algebra,
\eq{
\{\gamma^a,\,\gamma^b\}=2\,\eta^{ab}\,\mathds{1} 
}{eq:dharam1}
links gamma matrices $\gamma^a$ to spacetime geometry, in the simplest case given by the Minkowski metric $\eta^{ab}$ with signature $(-,+,\dots,+)$, thereby enabling Lorentz-covariant spinor equations like the Dirac equation, $(i\gamma^a\partial_a-m)\Psi=0$. 

If we were to summarize Carroll spinors in a single sentence, it would state that we replace the minus sign in the signature by zero and try to make sense of the implications for the gamma matrices and spinors. Of course, this is a very formalistic way of expressing the essence and does not do justice to the subject. So, the remainder of this work expands on this single sentence.

In Section \ref{sec:1}, we summarize key features of Carroll symmetries. In Section \ref{sec:2}, we explain in detail what Carroll spinors are. In Section \ref{sec:3}, we give a brief summary of applications where Carroll spinors have some role to play. The only part with new results is Subsection \ref{se:3.5} where we construct Carroll ELKO and the Appendices, while the rest is review material intended for the expected readership of this memorial volume, i.e., for researchers working on neutrino physics or ELKO.


\section{Carroll}\label{sec:1}

\textit{One of Dharam's side quests was the investigation of spacetime symmetries different from Poincar\'e \cite{Ahluwalia:2010zn} if only to point out that they might not differ from Poincar\'e after all \cite{Ahluwalia:2002wf}. So he appreciated the relation between (conformal) Carroll symmetries and asymptotic symmetries of flat space. I had the chance to introduce Dharam to my lower-dimensional Carrollian approach to flat space holography \cite{Bagchi:2012yk,Bagchi:2013lma,Bagchi:2013hja} during our encounter in 2013 towards the end of his visiting professorship in Campinas, Brazil.}

Carroll symmetries are spacetime symmetries for metrics with degenerate signature $(0,+,\dots,+)$ and arise as the vanishing speed of light limit of Poincar\'e symmetries \cite{Levy1965,SenGupta1966OnAA}, see Ref.~[\refcite{Bagchi:2025vri}] for a recent review. Carroll spacetimes are characterized by a degenerate Carroll metric $h$ and a Carroll vector $v$ in its kernel,
\eq{
h = h_{\mu\nu}\,\extd x^\mu \extd x^\nu\qquad\qquad v=v^\mu\,\partial_\mu\qquad\qquad h_{\mu\nu}\,v^\nu=0\,.
}{eq:dharam2}
The flat Carroll metric is obtained as $c\to 0$ limit of the Minkowski metric,
\eq{
\lim_{c\to 0}\big(-c^2\,\extd t^2+\delta_{ij}\,\extd x^i\extd x^j\big) = \delta_{ij}\,\extd x^i\extd x^j = h
}{eq:dharam3}
and the corresponding vector field (squared) as $c\to 0$ limit of the inverse Minkowski metric (rescaled by $c^2$),
\eq{
\lim_{c\to 0}\big(-\partial_t^2 + c^2 \delta^{ij}\partial_i\partial_j\big) = -\partial_t^2 = -v\otimes v\,.
}{eq:dharam4}
Since $h$ is degenerate it has no inverse and $-v\otimes v$ is instead playing the role of a ``metric with upper indices''.

Conformal Carroll symmetries are generated by vector fields $\xi$ that solve the conformal Carroll Killing equations,
\eq{
({\cal L}_\xi h)_{\mu\nu} = -2\lambda \,h_{\mu\nu}\qquad\qquad ({\cal L}_\xi v)^\mu = \lambda \,v^\mu\,.
}{eq:dharam5}
The proof by Duval, Gibbons, and Horvathy that conformal Carroll symmetries in $D$ spacetime dimensions are isomorphic to Bondi--van-der-Burgh--Metzner--Sachs (BMS) symmetries in $D+1$ spacetime dimensions \cite{Duval:2014uva} was a key insight for flat space holo\-graphy and is one of the main reasons for the exponential increase of Carroll-related research papers in recent years.


\section{Carroll spinors}\label{sec:2}

\textit{This memorial volume is the perfect opportunity to merge two of Dharam's favorite subjects, exotic spacetime symmetries and exotic spinors, manifested as Carroll spinors. While this topic is too recent to have featured in my discussions with Dharam in Zacatecas, Campinas, or via e-mail exchanges, I believe he would have enjoyed the construction reviewed in this Section.} 

Carroll fermions were introduced a few years ago \cite{Banerjee:2022ocj,Koutrolikos:2023evq,Bergshoeff:2023vfd}. Here, we review key aspects of Carroll spinors. To convey the main structure, we confine ourselves to 1+1 spacetime dimensions, except in the last Subsection.

This Section is organized as follows. In Subsection \ref{se:3.1}, we display the Carroll--Clifford algebra and provide an explicit representation thereof in terms of Carollian gamma matrices. In Subsection \ref{se:3.2}, we construct different types of Carrollian fermionic theories, dubbed ``electric'' and ``magnetic''. In Subsection \ref{se:3.3} we discuss charge conjugation and parity. In Subsection \ref{se:3.4}, we review the Carroll--Ising model as the simplest Carroll-invariant Lagrangian featuring spinors. Finally, in Subsection \ref{se:3.5}, we pose and answer the question of whether or not there are Carroll eigenspinors of the charge conjugation operator (Carroll ELKO).

\subsection{Carroll--Clifford algebra}\label{se:3.1}

The Carroll--Clifford algebra
\eq{
\{\gamma_a,\,\gamma_b\}=2\,h_{ab}\,\mathds{1}\qquad\qquad \{\gamma^a,\,\gamma^b\}=2\,V^{ab}\,\mathds{1}
}{eq:dharam6}
contains the flat Carroll metric $h_{ab}=\textrm{diag}(0,1)$ and its upper-index version, $V^{ab}=-v^av^b=\textrm{diag}(-1,0)$. As a consequence, the gamma matrices $\gamma_0$ and $\gamma^1$ are nilpotent, while $\gamma_1$ (and $\gamma^0$) square to (minus) unity. We can either adopt the homogeneous representation, where the nilpotent matrices are taken to be identically zero, or an inhomogeneous one where the whole set of matrices is non-trivial. A possible choice for the latter is\footnote{%
This choice is not unique; e.g., the Carroll--Clifford algebra can be inhomogeneously represented in terms of Pauli matrices, $\gamma^0=i\sigma_2$,  $\gamma_1=\sigma_1$, $\gamma_0=-\sigma_3-i\sigma_2$, and $\gamma^1=-i\sigma_3+\sigma_1$, see Appendix A of [\refcite{Grumiller:2024dql}]. Properties like $\mathcal{C}\mathcal{P}=1$ make this representation more akin to a Lorentzian one. 
\label{fn:1}} 
\eq{
\gamma_0=\begin{pmatrix}
    0 & a & \\ 0 & 0
\end{pmatrix} \qquad
\gamma_1=\begin{pmatrix}
    1 & 0 \\ 0 & -1
\end{pmatrix}, \qquad
\gamma^0=\begin{pmatrix}
    i & 0 & \\ 0 & -i
\end{pmatrix} \qquad
\gamma^1=\begin{pmatrix}
    0 & 0 \\ ia & 0
\end{pmatrix}
}{eq:dharam7}
where $a$ is an arbitrary real or imaginary parameter that we set to $+ 1$ without loss of generality. It is straightforward to see that this set of matrices is closed under transposition and complex conjugation, and there is a direct relation between upper and lower Carroll gamma matrices given by $\gamma^0=i\gamma_1$ and $\gamma^1 =-i {\gamma_0}^T$.

To make contact with Carrollian symmetries, and in analogy with the Lorentzian case, it is possible to define the ``upper-index'' commutator between the gamma matrices as
\eq{
\Sigma^{ab} = \frac{1}{4}\left[ \gamma^a,\, \gamma^b\right]
}{eq:dharam8}
and to show that the $\Sigma^{ab}$ span the Carroll algebra. In our two-dimensional case, the only non-vanishing component is $\Sigma^{01}$, which generates Carroll boosts. The four-dimensional case is summarized in \ref{app:A}. 

Equipped with the tools of this and the previous Subsection, in the next Subsection, we construct Carrollian actions involving spinors.

\subsection{Electric and magnetic fermions}\label{se:3.2}

One way we can construct a Carrollian action is by plugging a representation of the upper index gammas into the Dirac action
\eq{S=\int \extd^2 x \, \bar{\Psi} \big( i \gamma^a \partial_a - m) \Psi
}{eq:dharam9}
where $\Psi$ is a massive fermion field and $\bar{\Psi} = \Psi^{\dagger} \Lambda$ its adjoint. In the standard Lorentzian framework, one sets $\Lambda = \gamma^0$ to recover the Dirac adjoint. However, this identification does not necessarily hold in the Carrollian context. 

Following the analysis in [\refcite{Banerjee:2022ocj}], we consider Carroll transformations acting on spinors as $\Psi \to S(\Sigma)\Psi$, and seek a Hermitian matrix $\Lambda$ such that the adjoint spinor transforms as $\bar{\Psi} \to \bar{\Psi} S^{-1}(\Sigma)$, where $\Sigma$ are the generators defined in equation~\eqref{eq:dharam8}. Thus, it only amounts to finding a matrix $\Lambda$ that satisfies
\eq{{\Sigma^{ab}}^\dagger = - \Lambda \Sigma^{ab} \Lambda^{-1} \quad \Rightarrow \quad S(\Sigma)^{\dagger} = \Lambda S(\Sigma)^{-1} \Lambda^{-1}\,.
}{eq:dharam10}
For the representation \eqref{eq:dharam7}, the Carroll adjoint matrix can be chosen as
\eq{
\Lambda =\begin{pmatrix}
    0 & i & \\ -i & 0
\end{pmatrix}
}{eq:dharam11}
which satisfies $\Lambda^{-1}= \Lambda=\Lambda^{\dagger}$ and ${\gamma^{a}}^\dagger = \Lambda \gamma^{a} \Lambda^{-1}$. Using this last property, the Lagrangian defined in \eqref{eq:dharam9} can be proved to be hermitian. In terms of the spinor components $\Psi = (\psi_0, \psi_1)^T$, the Carroll spinor action is given by  
\eq{
\textrm{M-Carroll:}\quad S=i\,\int \extd^2 x \,\Big( \psi_0^\ast \dot{\psi}_1 + \psi_1^\ast \dot{\psi}_0 - \psi_0^\ast  \psi_0^\prime - m \psi_0^\ast \psi_1 + m \psi_1^\ast \psi_0 \Big)}{eq:dharam12}
where dot and prime denote time and space derivatives, respectively.

We see the appearance of terms involving both time and spatial derivatives, which looks more akin to that of Lorentzian fermions and leads to a non-vanishing Hamiltonian density.\cite{Bagchi:2022eui} However, the field $\psi_1$ acts as a Lagrange multiplier, enforcing the equation of motion $\dot{\psi}_0 = 0$, i.e., $\psi_0$ is non-dynamical. More generally, Carrollian theories with spatial derivatives in the action and constraints on time-derivatives of the fields are known as \textbf{magnetic}. 

Another approach is to construct a theory using lower-index gamma matrices instead of upper-index ones. This alternative also leads to a Carrollian theory, but with different characteristics. We consider the following Carroll-invariant terms: $\gamma_a V^{ab}\partial_b \Psi=-\gamma_0 \partial_0\Psi$ represents the only viable dynamical term and, regarding mass, the options are $m \bar{\Psi}\Psi$ and $i m \bar{\Psi} \gamma_0\Psi$. We choose the latter, as the former would lead to a trivial theory. Accordingly, we construct the action as follows:
\eq{
S=\int \extd^2 x \, \bar{\Psi} \gamma_0 \big(- \partial_0 - i m \big)  \Psi \,.
}{eq:dharam13}
Despite unusual factors of $i$, both terms are hermitian (in contrast to the upper gammas, the relation $\gamma_a^\dagger = -\Lambda \gamma_a \Lambda^{-1}$ holds). In components, the Carroll spinor action reads 
\eq{
\textrm{E-Carroll:}\quad S= \int \extd^2 x \, \Big( i\, \psi_1^\ast \dot{\psi}_1 - m \psi_1^\ast \psi_1 \Big) \, .
}{eq:dharam14}
This action has only time derivatives, which makes it ultra-local. More generally, Carrollian theories without spatial derivatives in the action are known as \textbf{electric}. 

The existence of electric and magnetic sectors is a typical feature of any theory with Carroll symmetries and can be understood as two different ways of performing the expansion in $c \to 0$ of a relativistic theory; see [\refcite{Bagchi:2025vri}] and Refs.~therein. These names stem from non-relativistic (Galilean) electromagnetism, where a similar dichotomy occurs \cite{LeBellac:1973unm}. For the next two Subsections, we focus mostly on magnetic fermionic actions \eqref{eq:dharam12}.

\subsection{Charge conjugation, parity and chirality}\label{se:3.3}

The unitary charge conjugation operator $\mathcal{C}$ acts on the fermionic field to produce its antiparticle counterpart by $\Psi \to \Psi^c = \mathcal{C} \, \bar{\Psi}^T$. Its matrix representation is defined by requiring it to leave the Dirac action invariant, which amounts to demanding
\eq{\mathcal{C}^{-1} \gamma^a \mathcal{C} = - \big(\gamma^a \big)^T
}{eq:dharam15}
together with $\mathcal{C}^{\dagger} \mathcal{C}=1$ and $\mathcal{C}^2 = \pm \mathds{1} $ (the sign depends on a representation choice). Considering the definition of the adjoint matrix in the Subsection above, and that our representation of upper-index gamma is purely imaginary, we choose the charge conjugation matrix as
\eq{
\mathcal{C}=\pm\Lambda\,. 
}{eq:dharam42}
We pick the minus sign in \eqref{eq:dharam42} to define Majorana spinors where the Majorana condition $\Psi := \Psi^c = \mathcal{C} \Lambda^T \Psi^\ast = \Psi^\ast$ enables a representation with real spinor field. 

A parity operator can be defined as acting on the fermion field generating a spatial coordinate inversion by $\Psi(t,x) \to \Psi^p = \mathcal{P} \, \Psi(t,-x)$. Requiring the Dirac equation to be invariant under this operation amounts to demanding $\mathcal{P}$ to commute with $\gamma^0$ and anticommute with $\gamma^1$, while preserving its projector nature, i.e., $\mathcal{P}^2=\mathds{1}$. We choose 
\eq{
\mathcal{P} = i \gamma^0
}{eq:dharam41}
which coincides with a standard choice in the Lorentzian case. However, there is a significant difference: the Carrollian bilinears $\bar{\Psi} \Psi$ and $\bar{\Psi} \gamma^\mu \Psi$ transform as a pseudoscalar and pseudovector, respectively, under parity, whereas in the Lorentzian setting they do as a scalar and vector. This is a general feature independent of the chosen representation: Lorentzian parity-odd and -even quantities are interchanged in the Carroll case.  

To construct parity-even bilinears, we introduce another gamma matrix $\gamma^\ast$ such that $(\gamma^\ast)^2=1$ and the objects $\bar{\Psi} \gamma^\ast \Psi$ and $\bar{\Psi} \gamma^\ast \gamma^\mu \Psi$ transform as a scalar and vector, respectively. A good choice is
\eq{
\gamma^\ast  =\begin{pmatrix}
    0 & 1 & \\ 1 & 0
\end{pmatrix}\,.
}{eq:dharam16}
It is worth mentioning that, in contrast with the Lorentzian setup, we cannot construct this additional gamma matrix as $\gamma^\ast\stackrel{?}{=} \gamma^0 \gamma^1$ in an upper-gamma Carroll representation, because it would square to zero. Our choice of $\gamma^\ast$ is useful for the purpose of flipping the sign of the bilinears under parity transformations, but it does not anticommute with the other two gamma matrices, and cannot be used to define chirality in any meaningful way.\footnote{%
To have a notion of chirality in our action \eqref{eq:dharam9}, it is necessary to identify two decoupled sectors of the spinor that transform independently under Carroll boosts. Since the latter are generated by the matrix $\Sigma$, a minimal requirement would be $[\Sigma, \gamma^\ast] = 0$. However, this only occurs if $\gamma^\ast$ anticommutes with all the other gamma matrices. It is easy to show that no such $\gamma^\ast\neq 0$ exists. \label{fn:2}} 

It is tempting, in the attempt to define a Carroll chiral theory, to consider the electric action given in \eqref{eq:dharam14}. From the Lagrangian perspective, it is clear that we obtain dynamics for a single degree of freedom (namely, $\psi_1$), which remains invariant under Carroll boosts. While this is indeed the case, as can be explicitly seen by considering the action of a Carroll boost $\delta_C \Psi = (x^1 \partial_0 - \Sigma_{01}) \Psi$ with $\Sigma_{01} = -\gamma_0/2$, it also turns out that the component $\psi_0$ transforms into the dynamical one. As a consequence, there is a mixing between components, implying that the best we can achieve in a Carroll theory is a single decoupled, boost-invariant chiral sector. This is consistent with the impossibility of using $\gamma^\ast$ to define a chiral projector.

\subsection{Carroll--Ising model}\label{se:3.4}

In this Subsection, we review a simple example of a Carroll conformal field theory in two dimensions (CCFT$_2$), namely the action of a massless magnetic Carroll fermion. We choose the inhomogeneous Majorana representation, in which spinors are real-valued quantities. This theory has been proven to be a free-field realization of the Carroll--Ising model at its critical point \cite{Yu:2022bcp,Hao:2022xhq}. The massless version of the magnetic Carroll action \eqref{eq:dharam12},
\eq{
\textrm{Carroll--Ising}:\quad  S=\int \extd^2 x\,  \big( \psi_0 \dot{\psi}_1 + \psi_1 \dot{\psi}_0 - \psi_0  \psi_0^\prime\big)
}{eq:dharam17}
yields the equations of motion
\eq{\dot{\psi}_0=0\qquad \qquad \dot{\psi}_1= \psi_0^\prime\,.
}{eq:dharam18}

Using these, the on-shell stress tensor $T^a{}_b$,
\eq{
T^0{}_0 = - T^1{}_1 = - \frac{1}{2} \psi_0 \psi_0^\prime \quad\qquad T^0{}_1 = - \frac{1}{2} \psi_1 \psi_0^\prime  - \frac{1}{2} \psi_0 \psi_1^\prime \qquad\quad T^1{}_0=0
}{eq:dharam19}
obeys the conformal Carroll Ward identities. The vanishing of the $T^1{}_0$ component is the boost Ward identity of the theory, which comes from demanding the action to be invariant under local Carroll boosts, whereas the vanishing of the trace $T^0{}_0 + T^1{}_1=0$ is the trace Ward identity required by Weyl invariance. As a result, there are only two independent components in the stress tensor, just like in a relativistic CFT in two dimensions. 

In the quantum theory, we promote the fields to operators and define $T(t,x) =\,\colon\!T^0{}_1\colon$ and $M(x) =\, \colon\!T^0{}_0\colon$ (the colons denote some notion of normal ordering for the fields). Note that, according to the equations of motion, $M$ depends solely on the spatial coordinate when evaluated on-shell, and satisfies the conformal Carroll Ward identity $\dot{T} = M^\prime$. This last identity\footnote{The same identity also arises in the context of asymptotically flat gravity in three dimensions, where it is known as BMS--Ward identity \cite{Bagchi:2015wna}.
\label{fn:holo}} shows that the time-dependent part of $T$ is fully determined by $M$ so that the full Carroll stress tensor is captured by two free functions of the spatial coordinate. 

On the cylinder or torus, it is convenient to expand both of them in modes: $L_n$ for the time-independent sector of $T$ and $M_n$ for the modes of $M$. We expect these operators to acquire central charges at the quantum level. We define the vacuum state $|0\rangle$ in the highest weight representation, satisfying
\eq{
L_n |0\rangle = 0 = M_m |0\rangle \qquad n,m \geq -1\,.
}{eq:dharam20}
Computing the OPE of the currents of the theory allows checking that these modes satisfy the CCFT$_2$ algebra (also known as BMS$_3$ algebra) \cite{Barnich:2006av,Barnich:2010eb,Bagchi:2010zz,Bagchi:2012yk,Campoleoni:2016vsh}
\begin{subequations}
    \label{eq:dharam21}
\begin{align}
    \big[L_n,L_m \big]&= (n-m) L_{n+m} + \frac{c_{\textrm{\tiny L}}}{12} n(n^2-1) \delta_{n+m,0}\\
    \big[L_n,M_m \big]&= (n-m) M_{n+m} + \frac{c_{\textrm{\tiny M}}}{12} n(n^2-1) \delta_{n+m,0}\\
    \big[M_n,M_m \big]&=0
\end{align}
\end{subequations}
where $c_{\textrm{\tiny L}}=1$ and $c_{\textrm{\tiny M}}=0$ for the Carroll--Ising model. In the context of three-dimensional gravity theories without a cosmological constant, the appearance of these generators reflects the asymptotic symmetries of flat space and suggests a dual description in terms of a CCFT$_2$ at its asymptotic boundary, albeit one with $c_{\textrm{\tiny L}}=0$ and $c_{\textrm{\tiny M}}\neq 0$ \cite{Bagchi:2010zz,Barnich:2010eb,Barnich:2012aw,Barnich:2012rz,Bagchi:2012yk,Bagchi:2012xr,Barnich:2012xq,Bagchi:2013lma,Duval:2014uva,Bagchi:2014iea,Bagchi:2015wna,Hartong:2015usd,Bagchi:2016bcd,Ciambelli:2018wre,Donnay:2022aba,Bagchi:2022emh,Donnay:2022wvx,Bagchi:2023fbj,Saha:2023hsl,Salzer:2023jqv,Saha:2023abr,Mason:2023mti,Chen:2023naw,Nguyen:2023vfz,Alday:2024yyj,Kraus:2024gso,Bagchi:2024efs,Bagchi:2024gnn,Ruzziconi:2024kzo,Chakrabortty:2024bvm,Fiorucci:2025twa,Hao:2025btl}.

This CCFT$_2$ algebra can also be obtained as the Carroll limit of the Lorentzian Ising model. We consider two copies of the Virasoro algebra 
\begin{subequations}
    \label{eq:dharam22}
\begin{align}
    \big[\ell_n,\ell_m \big]&= (n-m)\, \ell_{n+m} + \frac{c}{12}\, n(n^2-1)\, \delta_{n+m,0}\\
    \big[\bar{\ell}_n,\bar{\ell}_m \big]&= (n-m)\, \bar{\ell}_{n+m} + \frac{\bar{c}}{12}\, n(n^2-1) \,\delta_{n+m,0}
\end{align}
\end{subequations}
where $c=\bar{c}=\frac12$ are the central charges and $\ell_n$, $\bar{\ell}_n$ the generators of the holomorphic and antiholomorphic sectors of the CFT. 

To obtain the CCFT$_2$ algebra \eqref{eq:dharam21} as a limit, we consider a ``flipped'' vacuum state \cite{Bagchi:2020fpr} that is highest-weight for the holomorphic sector, $\ell_n |0\rangle=0$ for $n\geq -1$, but lowest-weight in the antiholomorphic one, $\bar{\ell}_{-n} |0\rangle=0$ for $n\geq -1$. This mixed condition on the vacuum state in the parent CFT yields the highest-weight vacuum in the limiting CCFT, with the CCFT generators related to the Virasoro generators by
\eq{
L_n := \ell_n - \bar{\ell}_{-n}\qquad\qquad M_n:= \epsilon( \ell_n + \bar{\ell}_{-n})\,.
}{eq:dharam23}
After taking the $\epsilon \to 0$ limit, both the algebra \eqref{eq:dharam21} and the highest weight conditions \eqref{eq:dharam20} are recovered. 

Had we used instead the standard CFT vacuum as starting point, the relations \eqref{eq:dharam23} would have yielded the CCFT central charges $c_{\textrm{\tiny L}} = c - \bar{c}$ and $c_{\textrm{\tiny M}} = \epsilon (c + \bar{c})=0$ and the so-called induced vacuum in the CCFT \cite{Barnich:2014kra,Oblak:2015sea}. However, starting with the flipped vacuum, upon exploiting the Virasoro algebra automorphism $\bar{\ell}_n\to-\bar{\ell}_{-n}$, $\bar c\to-\bar c$, we get another sign flip and the central charges of the limiting CCFT (see the discussion in Ref.~[\refcite{Aggarwal:2025hji}], especially Section 7.1 and Appendix B),
\eq{
\textrm{Carroll--Ising:}\quad c_{\textrm{\tiny L}} = c - (-\bar{c}) = \frac12+\frac12= 1\qquad\qquad c_{\textrm{\tiny M}} = \lim_{\epsilon\to 0}\epsilon (c + (-\bar{c})) = 0
}{eq:dharam40}
recover precisely the central charges of the Carroll--Ising model. 

Most of the statements above generalize to arbitrary CCFTs in two dimensions. In particular, the reviewed construction also works for all other CFTs as long as the flipped vacuum is used in the parent CFT. The limiting CCFT (with highest-weight vacuum) has a Virasoro central charge $c_{\textrm{\tiny L}}=c+\bar c$ that is the sum of the original Virasoro central charges and vanishing BMS central charge, $c_{\textrm{\tiny M}}=0$. This concludes our brief excursion into explicit models with Carroll spinors.

The example above shows that if one is careful with the choice of vacuum, it is possible to obtain non-trivial Carrollian results, even at the quantum level, as a limit from Lorentzian results. More generally, it suggests that Lorentzian structures have Carrollian counterparts. This motivates our search for the Carroll counterpart of ELKO, which we pursue in the next Subsection.

\subsection{Carrolleigenspinoren des Ladungskonjugationsoperators?}\label{se:3.5}

In this Subsection, we discuss the possibility of defining Carroll ELKO, i.e., Carrollian eigenspinors of the charge conjugation operator. Inspired by the original work in [\refcite{Ahluwalia:2004sz}], we define the operator as $C =\mathcal{C} \Lambda^T K$, where $\mathcal{C}$ is the charge conjugation matrix, $\Lambda$ the adjoint matrix, and $K$ denotes the action of complex conjugating the spinorial field. In the two-dimensional representation discussed in the previous Subsection \ref{se:3.3}, this operator simply reads $C=K$, where it can be seen that the eigenstates are Majorana spinors with eigenvalue $+1$. Since ELKOs were originally defined in a four-dimensional setting, we now turn our attention to that case. A detailed generalization of some of the results discussed in the previous Subsections to four dimensions can be found in \ref{app:A}.

A representation for the four-dimensional ``lower'' Carroll--Clifford algebra that allows for the construction of non-trivial ELKOs is provided by
\eq{
\gamma_0= \frac{1}{2}\begin{pmatrix}
    i \, \mathds{1} & \, \mathds{1} & \\  \, \mathds{1} & -i \, \mathds{1}
\end{pmatrix} \qquad\qquad
\gamma_i=\begin{pmatrix}
    \boldsymbol{0} & i \, \sigma_i \\ -i \, \sigma_i & \boldsymbol{0}
\end{pmatrix}
}{eq:dharam24}
where $\boldsymbol{0}$ and $\mathds{1}$ denote the $2 \times2$ zero block matrix and identity, respectively, and the $i$ indices correspond to the three spatial directions. For this representation, the Carroll adjoint matrix can be chosen as
\eq{
\Lambda= \begin{pmatrix}
    \mathds{1} & \boldsymbol{0} & \\ \, \boldsymbol{0}& -\mathds{1}
\end{pmatrix}
}{eq:dharam25}
which satisfies $\Lambda= \Lambda^{-1}=\Lambda^\dagger$ and $\gamma_a^{\dagger}=-\Lambda \gamma_a \Lambda^{-1}$.

As stated at the end of Subsection \ref{se:3.3}, for Carrollian fermions in 2D it is not possible to define a notion of chirality. The same occurs in the four-dimensional case, where the Carroll symmetry group does not correspond to the usual Lorentz group SO$(3,1)$, and therefore cannot be represented as SU$(2)_L \times$ SU$(2)_R$. Our goal is to define ELKOs solely as the eigenspinors of the charge conjugation operator. To do that, and inspired by the Lorentzian case, we plan to take advantage of the 2$\times$2 block structure of the chosen representation to split the four-dimensional spinor in terms of two two-component spinors $\Psi = (\phi_0, \phi_1)^T$ with $\phi_0, \phi_1 \in \mathbb{C}^2$. In Subsection \ref{se:3.2} it was shown that the electric fermion theory contains no spatial derivatives and, therefore, no spatial gamma matrices. Consequently, as a basis for constructing Carroll ELKOs, we define the action of the charge conjugation as 
\eq{
\mathcal{C}^{-1}\gamma_0 \mathcal{C} = - (\gamma_0)^T
}{eq:dharam25a}
while demanding unitarity $\mathcal{C}^\dag \mathcal{C} = 1$. For the given representation \eqref{eq:dharam24} these conditions are obeyed  by
\eq{ 
\mathcal{C}=\begin{pmatrix}
    \boldsymbol{0} & \mathcal{A} & \\ -\mathcal{A} \, & \, \boldsymbol{0}\end{pmatrix}
}{eq:dharam26}
with $\mathcal{A}$, a $2\times 2$ matrix that satisfies $\mathcal{A}^\dag \mathcal{A} = \mathds{1}$. Confining ourselves to real eigenvalues, eigenstates of the charge conjugation operator $C$ with eigenvalues $\pm 1$  can be found in terms of the components
\eq{
\lambda^{\pm} = \begin{pmatrix}
    \phi_0 \\   \mp\ \mathcal{A} \ \phi_0^\ast
\end{pmatrix}
\qquad \text{and} \qquad
\rho^\pm = \begin{pmatrix}
    \mp \mathcal{A} \ \phi_1^\ast \\  \phi_1
\end{pmatrix}
}{eq:dharam27}
with $\mathcal{A} \mathcal{A}^\ast = \mathds{1}$ and $\mathcal{A}^\dag\mathcal{A} = \mathds{1}$, such that
\eq{
C \lambda^\pm= \pm \lambda^\pm \qquad\qquad C \rho^\pm= \pm \rho^\pm\, .
}{eq:dharam28}

This definition of Carroll ELKOs is, however, not a Carroll invariant one, given that boosts inevitably mix the two-dimensional components of the Carroll spinor -- as can be seen from equations \eqref{eq:dharamA5a} and \eqref{eq:dharamA5b} -- and expressions such as \eqref{eq:dharam27} do not close under the action of the full Carroll group. 

Instead, we find a proper subgroup that leaves the Carroll ELKOs invariant. The closure conditions for an infinitesimal transformation with parameter $\omega$ can be expressed as
\eq{
\delta_\omega \phi_1 \big|_{\lambda^{\pm}} = \mp \mathcal{A} \, \delta_\omega \phi^\ast_0 \big|_{\lambda^{\pm}} \qquad \text{and} \qquad \delta_\omega \phi_0 \big|_{\rho^{\pm}} = \mp \mathcal{A} \, \delta_\omega \phi^\ast_1 \big|_{\rho^{\pm}}
}{eq:dharam29}
which yields the requirements $\sigma_i = \mathcal{A} \, \sigma_i^\ast \, \mathcal{A}^\ast$ for Carroll boosts in the $i$-direction and $\sigma_i = -\mathcal{A} \, \sigma_i^\ast \, \mathcal{A}^\ast$ for spatial rotations around the $i$-axis. There is no $\mathcal{A}$ that can fulfill all these requirements simultaneously. 

We can choose, for instance, $\mathcal{A}= \sigma_1$, which specifies a preferred direction\footnote{%
Alternative choices for $\mathcal{A}$ include $\mathds{1}$ and $\sigma_3$, which correspond to the two remaining spatial directions that could be preferred. Any direction in three-dimensional space can be represented as a linear combination of these matrices.
}. Then, only Carroll boosts in the directions $1$ and $2$, and spatial rotations around the axis $3$ preserve the structure of the Carroll ELKOs. Notably, something similar happens with the Lorentzian ELKOs, which are also not invariant under the whole homogeneous Lorentz group but rather the smaller 4-parameter subgroup SIM$(2)$ \cite{Ahluwalia:2008xi,Ahluwalia:2009rh,Boehmer:2010ma,Ahluwalia:2010zn}, the group of transformations that preserve the direction of a null vector in Minkowski spacetime. In the Carroll case it can be shown that the generators of the three invariant transformations we found are compatible with a Carrollian In\"on\"u--Wigner contraction of the SIM$(2)$ algebra. A detailed computation can be found in \ref{app:B}.

There are numerous open questions concerning Carroll ELKOs, some of which we list in the final Section.


\section{Applications involving Carroll spinors}\label{sec:3} 

\textit{In his research, Dharam insisted on a phenomenological perspective, especially in mostly theory-driven fields like quantum gravity \cite{Ahluwalia:1999aj}. Therefore, it is apposite to conclude with a phenomenological Section on applications involving Carroll spinors, followed by a todo list regarding Carroll ELKO.}

What are Carroll spinors good for? In this last Section, we address this question, starting from condensed matter applications, continuing with gravitational applications, and finishing with superstring applications. (We do not review many of the bosonic Carroll applications like gravitational pp-waves, BKL physics, fractons, swiftons or Carroll black holes, even though several of them may have generalizations featuring Carroll spinors; see the recent review [\refcite{Bagchi:2025vri}] for such applications.)

One key aspect of Carroll symmetries in position space is that the lightcone closes up. Conversely, in momentum space the lightcone flattens and one obtains a flat band structure. Approximately such structures are observed in certain condensed matter systems, such as magic angle bilayer graphene \cite{Bagchi:2022eui,Ara:2024fbr}. Thus, whenever a flat band structure is a good approximation, the system behaves approximately Carrollian.

A somewhat related set of applications is near horizon soft hair \cite{Hawking:2016msc}. ``Soft'' refers to excitations without energy, which can be rephrased as stating that the Hamiltonian is independent from the momentum, i.e., a flat band structure. Indeed, soft hair excitations were constructed explicitly in 2+1 dimensions \cite{Donnay:2015abr,Afshar:2016wfy} and shown to generate BMS$_3$ symmetries \cite{Afshar:2016kjj}, which are isomorphic to two-dimensional conformal Carroll symmetries \cite{Bagchi:2010zz,Duval:2014uva}. This story generalizes to four spacetime dimensions \cite{Donnay:2015abr,Grumiller:2019fmp}. While so far, the discussion was focused on bosonic excitations, we expect also fermionic excitations to be in line with (super-)conformal Carroll symmetries, and hence Carroll spinors should feature in near horizon soft hair.

Similarly, in the context of flat space holography the focus so far was mostly on bosonic excitations (see, however, [\refcite{Bagchi:2022owq}]). If we consider flat space as a vanishing cosmological constant limit of AdS, it is suggestive that one could perform a similar limit within the AdS/CFT correspondence. The latter, coming from superstring theory, inherently involves spinor fields so that also its flat space limit must contain spinors. This makes it necessary to include Carroll spinors in any complete description of flat space holography. 

Finally, the tensionless limit of superstrings also leads to Carrollian structures and requires Carroll spinors \cite{Bagchi:2016yyf,Chen:2023esw}. While this is a very-high-energy application (at the Hagedorn scale), for a complete description of the early Universe understanding tensionless superstrings may be required. For further literature on Carroll spinors and some applications thereof see, e.g., Refs.~[\refcite{Mele:2023lhp,Ekiz:2025hdn,Bergshoeff:2024ytq,deBoer:2023fnj,Zheng:2025cuw,Chen:2025gaz,Grumiller:2024dql,Bagchi:2025vri}].

We conclude with a list of comments and open questions regarding Carroll ELKO:
\begin{itemize}
    \item A key property of any Carrollian theory is that the Dirac--Schwinger condition on the energy density, $[\mathcal{E}(x),\mathcal{E}(x^\prime)]\sim (P^i(x)+P^i(x^\prime))\partial_i\delta(x-x^\prime)$ simplifies to the Bunster--Henneaux condition\cite{Teitelboim:1978uc,Henneaux:1979vn,Bunster:2012hm,Henneaux:2021yzg} 
    \eq{
    [\mathcal{E}(x),\mathcal{E}(x^\prime)] = 0\,.
    }{eq:BunsterHenneaux}
    In the classical theory, the commutators above are replaced by Poisson brackets. The Bunster--Henneaux condition \eqref{eq:BunsterHenneaux} provides a rigid cross-check for interacting Carrollian theories (such as swiftons\cite{Ecker:2024czx}) and could be a key consistency check when tinkering with Carroll ELKO actions. 
    \item Our Carroll ELKO discussion involved some choices: representations of the Carroll gamma matrices (including the choice of the rank of the nilpotent ones), electric vs.~magnetic Carroll spinors, preferred symmetry axis, etc. We have not made any attempt to be exhaustive concerning these choices or to classify which of them are merely conventional and which of them have a more profound impact. It could be rewarding to do this. 
    \item Some details of our Carroll ELKO construction remain puzzling. For example, one might expect that the Carroll contraction of SIM$(2)$ yields the Carroll ELKO symmetry algebra we have found in Subsection \ref{se:3.5} but this is not quite the case as there is an extra central term in such a contraction. Is this central term of any significance?
    \item Finally, any question addressed in the context of ELKO (see the review [\refcite{Ahluwalia:2022ttu}] and Refs.~therein) could in principle be addressed for Carroll ELKO. However, we should warn the reader that, currently, we do not envisage any specific physics application of Carroll ELKO.
\end{itemize}


\section*{Acknowledgments}

\textit{I am grateful to Julio M.~Hoff da Silva and Cheng-Yang Lee for the invitation to contribute to this Special Issue in memory of Dharam Ahluwalia.}

We thank Arjun Bagchi, Aritra Banerjee, Rudranil Basu, Eric Bergshoeff, Andrea Campoleoni, Sudipta Dutta, Florian Ecker, Andrea Fontanella, Jan Rosseel, and Mohanna Shams Nejati for discussions/collaboration on Carroll spinors.

This work was supported by the Austrian Science Fund (FWF) [Grants DOI: \href{https://www.fwf.ac.at/en/research-radar/10.55776/P33789}{10.55776/P33789}, \href{https://www.fwf.ac.at/en/research-radar/10.55776/P36619}{10.55776/P36619}, \href{https://www.fwf.ac.at/en/research-radar/10.55776/PAT1871425}{10.55776/PAT1871425}]. 

Lea Mele is a FRIA grantee of the Fund for Scientific Research — FNRS, Belgium. The work of Lea Mele was supported by FNRS under Grant FC.55077.

\section*{ORCID}

\noindent Daniel Grumiller - \url{https://orcid.org/0000-0001-7980-5394}

\noindent Lea Mele - \url{https://orcid.org/0009-0000-7672-883X}

\noindent Luciano Montecchio - \url{https://orcid.org/0000-0003-2996-069X}


\appendix

\section{Carroll spinors in four dimensions}\label{app:A}

In this Appendix, we provide the analogues for four dimensions of the results of the Subsections \ref{se:3.1}-\ref{se:3.2} that will be useful for our analysis of Carroll ELKO spinors. The results presented in this Appendix are mainly drawn from [\refcite{Bagchi:2022eui}] and [\refcite{Stakenborg:2023bmw}] \footnote{An analysis of Carroll fermions in four dimensions obtained through a Carroll limit can be found in [\refcite{Bergshoeff:2023vfd}], as well as in [\refcite{Koutrolikos:2023evq}].}.

The four-dimensional Carroll--Clifford algebra is
\eq{
\{\gamma_a,\,\gamma_b\}=2\,h_{ab}\,\mathds{1}\qquad\qquad \{\gamma^a,\,\gamma^b\}=2\,V^{ab}\,\mathds{1},
}{eq:dharamA1}
with the flat Carroll metric $h_{ab}=\textrm{diag}(0,1,1,1)$ and its upper-index version, $V^{ab}=-v^av^b=\textrm{diag}(-1,0,0,0)$.

We define the ``lower-index" commutator between the gamma-matrices as
\eq{
\Sigma_{a b} = \frac{1}{4}\left[ \gamma_a,\, \gamma_b\right]
}{eq:dharamA2}
which span the homogeneous Carroll algebra
\eq{ 
\begin{aligned}
& {\left[\Sigma_{0 i}, \Sigma_{0 j}\right]=0 \quad\left[\Sigma_{0 i}, \Sigma_{j k}\right]=\delta_{i j} \Sigma_{0 k}-\delta_{i k} \Sigma_{0 j}} \\
& {\left[\Sigma_{i j}, \Sigma_{k l}\right]=\delta_{i l} \Sigma_{j k}-\delta_{i k} \Sigma_{j l}+\delta_{j k} \Sigma_{i l}-\delta_{j l} \Sigma_{i k}}
\end{aligned}
}{eq:dharamA3}
where $\Sigma_{0i}$ is the generator for the Carroll boosts and $\Sigma_{ij}$ the one for the spatial rotations. Under infinitesimal Carroll transformations (with parameters $\omega^{ab} = - \omega^{ba}$), a Dirac spinor $\Psi$ transforms as
\eq
{
\delta \Psi = \omega^{a b} x_b \partial_a \Psi - \frac{1}{2}\omega^{ab}\Sigma_{ab}\Psi.
}{eq:dharamA4}
 A possible representation for the lower gamma matrices is given in \eqref{eq:dharam24}, with the Carroll adjoint matrix \eqref{eq:dharam25}. In this case, the Carroll transformation for the spinor $\Psi$ expressed in terms of its two-components spinors $\phi_0$ and $\phi_1$ takes the following form
\begin{subequations}
\begin{align}
\delta{\phi_0} &= \omega^{a b} x_b \partial_a \phi_0 + \frac{i}{4}\omega^{0i} \sigma_i \phi_0 + \frac{1}{4}\omega^{0i}\sigma_i \phi_1 - \frac{1}{8}\omega^{ij}\left[\sigma_i, \sigma_j\right] \phi_0  \label{eq:dharamA5a}\\
\delta{\phi_1} &= \omega^{a b} x_b \partial_a \phi_1 - \frac{i}{4}\omega^{0i} \sigma_i \phi_1 + \frac{1}{4}\omega^{0i}\sigma_i \phi_0 - \frac{1}{8}\omega^{ij}\left[\sigma_i, \sigma_j\right] \phi_1\,.
\label{eq:dharamA5b}
\end{align}
\end{subequations}

Even though it is not required for the definition of ELKO, for the sake of completeness, we introduce an electric action invariant under the Carroll transformations \eqref{eq:dharamA4}
\eq{
S=\int \, \extd^4 x \, \bar{\Psi} \gamma_0 \big(- \partial_0 - i m \big)  \Psi
}{eq:dharamA6}
where $\bar{\Psi} = \Psi^{\dag} \Lambda$. As in Subsection \ref{se:3.2} for the two-dimensional electric action, we choose the mass term $m \bar{\Psi}\gamma_0 \Psi$, and did not include the $i$-factor in the kinetic term. 

Regarding the upper-index gamma matrices, one can also define the “upper-index” commutator as
\eq{
\Sigma^{ab}= \frac{1}{4}\left[ \gamma^a, \gamma^b \right].
}{eq:dharamA7}
However, with this definition, one has that the rotations generator $\Sigma^{ij}$ becomes zero. With the former result and $V^{ab}=-v^av^b=\textrm{diag}(-1,0,0,0)$, this implies that the commutation relations between the generators vanish
\eq{
\left[\Sigma^{a b}, \Sigma^{c d}\right]=-\Sigma^{a d} v^b v^c+\Sigma^{b d} v^a v^c-\Sigma^{b c} v^a v^d+ \Sigma^{a c} v^b v^d=0
}{eq:dharamA8}
meaning that the structure does not close into the Carroll algebra. Nevertheless, a Carroll-boost-invariant Lagrangian can still be formulated for the theory with upper-index gamma matrices. An action invariant under the Carroll boosts transformations
\eq{
\delta \Psi = \omega_{0i}x^i \partial_0 \Psi - \omega_{0i}\Sigma^{0i}\Psi,
}{eq:dharamA9}
is given by
\eq{
S = \int \, d^4x \, \bar{\Psi}\left( i \gamma^a\partial_a - m  \right)\Psi 
}{eq:dharamA10}
with $\bar{\Psi} =  \Psi^\dag \Lambda$. A possible representation for the upper-index Clifford algebra is
\eq{
\gamma^0=\left(\begin{array}{cc}
i \mathds{1} & \boldsymbol{0} \\
\boldsymbol{0} & -i \mathds{1}
\end{array}\right) \qquad\qquad \gamma^i=\left(\begin{array}{cc}
\boldsymbol{0} & \boldsymbol{0} \\
-\sigma^i & \boldsymbol{0}
\end{array}\right)\,
}{eq:dharamA11}
with $\Lambda$ that can be chosen as
\eq{
\Lambda = \left(\begin{array}{cc}
    \boldsymbol{0} &  \mathds{1}\\
     \mathds{1}& \boldsymbol{0}
\end{array}
\right)\quad \text{ with } \gamma^{ a \dag  } = \Lambda \gamma^a \Lambda^{-1}\,.
}{eq:dharamA12}
Using this representation, the Carroll transformation \eqref{eq:dharamA9} in terms of the two two-components spinors $\phi_0$ and $\phi_1$ becomes
\begin{subequations}
\begin{align}
\delta{\phi_0} &=\omega_{0i}x^i \partial_0 \phi_0  \label{eq:dharamA13a}\\
\delta{\phi_1} &= \omega_{0i}x^i \partial_0 \phi_1 - \frac{i}{2} \omega_{0i} \sigma_i \phi_0\,.
\label{eq:dharamA13b}
\end{align}
\end{subequations}
In [\refcite{Bergshoeff:2023vfd}], a Carrollian magnetic theory for a Dirac field invariant under the full homogeneous Carroll group is obtained through a Carroll limit. The resulting Carroll transformations show that the two two-components spinors form a reducible but indecomposable representation of the homogeneous Carroll group, a feature also present in other Carroll magnetic theories such as the Carrollian scalar field \cite{Bergshoeff:2024ytq}.

\section{Carroll contraction of SIM$(2)$}\label{app:B}

In this Appendix, we show how to perform a Carroll contraction of the SIM$(2)$ algebra. We start by considering the $3+1$ dimensional homogeneous Carroll algebra
\eq{
\left[J_i,J_j \right]= i \epsilon_{ijk} J_k\quad \qquad \left[J_i,C_j \right]= i \epsilon_{ijk} C_k\quad \qquad \left[C_i,C_j \right]=0
}{eq:dharamB1}
where the $J_i$ and the $C_i$ are the generators of the spatial rotations and Carroll boosts, respectively. The generators $(C_1,C_2,J_3)$ span the subalgebra ISO$(2)$ of \eqref{eq:dharamB1},
\eq{
\left[J_3,C_1 \right]= i C_2\qquad \qquad \left[J_3,C_2 \right]= -i C_1
}{eq:dharamB2}
with the possibility of adding $C_3$ as a central generator that commutes with the rest and does not appear on the r.h.s.~of any commutator. We show next that these same commutators arise as a Carroll contraction of the SIM$(2)$ algebra, the latter being a four-parameter subgroup of the homogeneous Lorentz group preserving a fixed light-like direction. The SIM$(2)$ algebra is given by \cite{Cohen:2006ky,Ahluwalia:2010zn}
\begin{subequations}
\label{eq:dharamB3}
\begin{align}
\left[T_1,J_3\right]&= -i \, T_2 & \left[T_1,B_3\right]&= i \, T_1 & \left[T_1,T_2\right] &= 0\\
\left[T_2,J_3\right]&= i \, T_1 & \left[T_2,B_3\right] &= i \, T_2 &  \left[J_3,B_3\right] &= 0
\end{align}
\end{subequations}
where $J_3$ and $B_3$ represent a rotation and a Lorentz boost in the 3--axis and the $T$ generators are defined as $T_1 = B_1 +J_2$ and $T_2=B_2 - J_1$. We now perform a rescaling in powers of the parameter $c$, which represents the speed of light, 
\eq{
J_i \to J_i\qquad \qquad B_i \to C_i = c \, B_i\qquad \qquad T_{1,2} \to T'_{1,2}= c \, T_{1,2}
}{eq:dharamB4}
and by taking the Carroll limit (i.e., $c \to 0$ limit, leaving $C_i$ and $T'_{1,2}$ constant), we obtain the identifications $T'_1=C_1$ and $T'_2=C_2$, together with the commutation relations \eqref{eq:dharamB2} and $C_3$ as a central term that commutes with the rest. The non-abelian subgroup ISO$(2)$ corresponds precisely to the symmetry transformations proven to leave the Carroll ELKOs invariant.



\providecommand{\href}[2]{#2}\begingroup\raggedright\endgroup

\end{document}